\documentclass[10ft,conference]{IEEETran}
\usepackage{amsfonts}
\usepackage{amsmath,epsfig,amssymb}

\usepackage{txfonts}
\usepackage{eucal,times,latexsym}
\usepackage{color}
\usepackage{colortbl}
\usepackage{cite}

\IEEEoverridecommandlockouts

\begin{document}

\title{\huge Efficient ML Decoding for Quantum Convolutional Codes
}
\author{
Peiyu Tan and Jing Li (Tiffany)\\
Electrical and Computer Engineering Department, 
Lehigh University, Bethlehem, 18015 \\
\{pet3,jingli\}@ece.lehigh.edu\vspace{-0.5cm} }

\maketitle
\vspace{-2.4cm}

\begin{abstract}

A novel decoding algorithm is developed for general quantum convolutional
codes. Exploiting useful ideas from classical coding theory, the new decoder introduces two innovations that drastically reduce the decoding complexity compared to  the existing quantum Viterbi decoder. First, the new decoder uses an efficient linear-circuits-based mechanism to map a syndrome to a candidate vector, whereas the existing algorithm relies on a non-trivial lookup table. Second, the new algorithm is cleverly engineered such that only one run of the Viterbi algorithm suffices to locate the most-likely error pattern, whereas the existing algorithm must run the Viterbi algorithm many times. The efficiency of the proposed algorithm allows us to simulate and present the first performance curve of a general quantum convolutional code.


\end{abstract}

\footnotetext{Supported by NSF under Grants No. CCF-0635199, CCF-0829888 and CMMI-0928092.}

\section{Introduction}
\label{sec:introduction}


We consider quantum error correction coding (QECC) models  where quantum bits (qubits) and quantum errors are both modeled as linear combination of Pauli matrices ${\mathbf{I, X, Y, Z}}$. An [$n,k$] (block) stabilizer code\footnote{We use [...] to denote quantum stabilizer codes, (...) to denote classical binary codes, and (...)$_q$ to denote classical quaternary codes.} is defined as a codeword space $\mathcal {C}$ which is \emph{non-trivially} stabilized by a subgroup $\mathcal {S}$ of $\mathbb{G}_n$, where $\mathbb{G}_n$ is length-$n$ tensor product of Pauli-operators ${\bf I}$, ${\bf X}$, ${\bf Y}$, ${\bf Z}$. By definition,  the independent generators of $\mathcal {S}$ commute to each other, and $-{\bf I} \notin \mathcal {S}$.

The dichotomy between the unobservable state of a qubit and the observations we can make, lies at the heart of quantum computation including QECC.  In classic error correction, output from the channel is observed, measured and used directly in the decoding process. Observation in quantum mechanics not only destroys the quantum state and makes restoration impossible, but the measurement does not tell the original quantum state either. The measured outcome is always one of the two basic states, whose probability of occurring is proportional to the power projection of the original state on these two bases. For this reason, to decode a stabilizer code typically takes three steps:  measuring the syndrome, identifying the error pattern based on the
syndrome, and applying a corrective operation to reverse the error.
The syndrome diagnosis and the error reversal are well-established quantum mechanical procedures, and operate the same way regardless of the stabilizer configuration. Identifying error patterns, a process analogous to syndrome-decoding of classical codes, is most challenging. Not only are the techniques used here specific to the code structure, but they generally involve tedious (exhaustive) search in the Hilbert space.

Compared to stabilizer block codes, stabilizer convolutional codes
enjoy the flexibility in codeword length and
online encoding. The first comprehensive description of the general formalism of quantum
convolutional codes appeared in \cite{bib:QCCfundamentals}, and a variety of quantum convolutional codes has since been proposed \cite{bib:QC_ISIT}-\cite{bib:TailbitingCC}. These studies focus on the code construction (e.g. based on classical Reed-Solomon codes, classical low-density parity check codes, and concatenation of classical convolutional codes), and do not usually discuss feasible decoding algorithms.
The first decoder for quantum convolutional codes appeared in
\cite{bib:QCCfundamentals}, which developed a quantum Viterbi algorithm (QVA)   
close in spirit to the classical Viterbi algorithm. 
Simplification is made to this QVA
 by  attacking only a {\it single} unitary error per block (and ignoring all the other possible error events) \cite{bib:Forney,bib:TailbitingCC}.
For quantum convolutional codes constructed from quasi-cyclic sparse matrices, a different type of decoder based on pipeline message passing is developed \cite{bib:QC_ISIT}.
Since the decoder in \cite{bib:Forney,bib:TailbitingCC} tackles only single errors, and that in \cite{bib:QC_ISIT} performs well only on sparse convolutional codes, the QVA in \cite{bib:QCCfundamentals} is by far the only general and ML decoding method available for stabilizer convolutional codes. However,
it requires a very high complexity, caused by a very large lookup table and many rounds of the conventional Viterbi algorithm.
The high complexity and the induced long delay make this algorithm very challenging to implement (and it is for this practicality issue that \cite{bib:Forney,bib:TailbitingCC} proposed to simplify it at the cost of degraded performance).


The contribution of this paper is the development of a new decoding
algorithm  which works for a general quantum convolutional code and
which is drastically simpler than the existing one. Not using lookup
tables of any kind, the proposed algorithm exploits a simple
linear-sequential-circuits based mechanism  to map a length-$(n-k)$
(binary) syndrome to {\it a} length-$n$ (quaternary) candidate
vector. This candidate vector is subsequently fed to a conventional
Viterbi decoder to help identify the mostly likely error pattern.
The entire process is so cleverly engineered that a {\it single}
candidate vector -- any one among the pool of $2^{(n+k)}$ possible
candidate vectors -- suffices to locate {\it the} error pattern in
the ML sense.
The high efficiency of this algorithm enables us to, for the first time in quantum coding literature, simulate and present qubit error rate
performance curves for a general quantum convolutional code.


\section{Quantum Convolutional Codes}
\label{sec:background}

\subsection{Definition and Representations}

A stabilizer convolutional code may be viewed as an infinite version of
a stabilizer block codes with repeated structure. The stabilizer group, $\mathcal{S}$, for an [$n,k,m$] stabilizer convolutional code is given by \cite{bib:QCCfundamentals}:
\begin{equation}\label{equ:def_qcc}
    \mathcal{S} = sp\{M_{j,i} = I^{\otimes jn}\otimes M_{0,i},  1\leq i \leq n-k, 0 \leq
    j\},
\end{equation}
where $m$ is the so-called memory parameter, $M_{0,i} \in G_{n+mn} = sp\{{\bf I, X, Y, Z}
\}^{\otimes (n+mn)}$. \footnote{$G_{n+m\times n}$ in
\cite{bib:QCCfundamentals} was written as $G_{n+m}$, which is
either mistaken or a typo.}, and $M_{j,i}$'s are required to be
independent and to commute with each other. Like classical convolutional codes, quantum convolutional codes also have flexible and adjustable codeword lengths.

\begin{figure}[htbf]
  \vspace{-0.4cm}
\centerline{\includegraphics[width=3.3in]{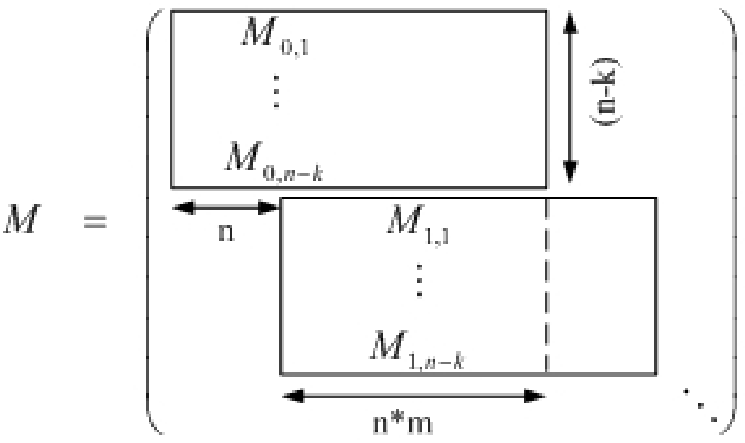}}
  \label{fig:semi_infinite}
  \vspace{-0.3cm}
\caption{Semi-infinite Matrix $M$ of a quantum convolutional code.}
\end{figure}

The structure of the stabilizer group generators can also be characterized
by a semi-infinite matrix $M$. Similar to the stabilizer block
codes, each line in $M$ represents a generator (some $M_{j,i}$),
and each column represents a qubit. As illustrated in Fig. \ref{fig:semi_infinite}, $M$ has a
block-band structure, and two neighboring blocks
of generators overlap by $mn$ qubits, representing the ``actual memory'' of the quantum convolutional code.

Similar to stabilizer block codes, we can label the
polynomial generators consisting of Pauli elements by $\mathbb{F}_2$
polynomial vectors in $D$-domain. The commutation property of stabilizers
translates to:\vspace{-0.3cm}
\begin{equation}\label{equ:symplectic_QCC}
    P(D) Q(\frac{1}{D})^T + Q(\frac{1}{D}) P(D)^T = 0,
\end{equation}
where $D$ stands for the delay element, similar to that in classical convolutional codes, and  $(P(D) | Q(D))$ denotes the $\mathbb{F}_2$ polynomial vectors
of a quantum stabilizer code.

\subsection{Existing Decoding for Quantum Convolutional Codes}

The quantum Viterbi algorithm discussed in \cite{bib:QCCfundamentals} is the only known general decoder for quantum convolutional codes. A syndrome
version of the classical Viterbi decoder, the QVA operates over
Pauli-operators (non-binary input).
It starts with a given list of syndromes, and searches for
 the candidate vectors, which are analogous to the ``received sequence (codeword)'' in classical Viterbi decoding.
The task is accomplished through table lookup, where the decoder examines the syndromes block by block.
Because of the
convolutional nature, the output of two neighboring syndrome blocks
overlap by $nm$ qubits, and the table lookup procedure must compare and ensure the overlapping part of the candidate vectors match. The result is a very long list of candidate vectors all of which correspond to the given syndromes.
Every one of these candidate vectors will then be fed into a classical Viterbi algorithm, and among them,  the most-likely sequence, namely, the error pattern, will be identified. The overall complexity is rather prohibitive \cite{bib:QCCfundamentals}.

To avoid such a high complexity and still be able to peek into the code's
performance, \cite{bib:TailbitingCC} proposed a simplified decoder that tackles only the most frequent (but not necessarily the most detrimental) error patterns, namely, no more than a single error per block.
As such, the algorithm is able to inspect only two
consecutive blocks at a time, and hence drastically reduces the size of the lookup table to only 9 terms, at the cost of a compromised decoder performance.


\section{Proposed Decoding Algorithm}
\label{sec:NewConstruction}

\subsection{Syndrome Decoding in Classical Coding}
The proposed decoding algorithm roots to the classical syndrome
decoding approach for linear codes.
 In essence, syndrome decoding is minimum
distance decoding, which is equivalent to the maximum likelihood
decoding if the channel is discrete memoryless with error probability
strictly less than 0.5. The linearity of a code allows the code space to be arranged in a special way, termed the {\it standard array} in coding jargon. For an ($n, k, d$) classical linear block code, the standard array groups all the $2^n$ $n$-bit vectors into $2^k$ columns and $2^{n-k}$ rows, such that (i) the vectors in the same row corresponding to the same syndrome, (ii) the first row consists of the set of valid codewords (starting with
the all-zero codeword), and (iii) the first column consists of all the {\it coset leaders}, i.e.
correctable error vectors with Hamming weights lower than
$\lfloor (d-1)/2 \rfloor$. Each row of the standard array is called a
\emph{coset}.
In concept, the function of a  syndrome decoder is  to identify the coset leader, i.e. the minimum-weight vector, corresponding to a given
syndrome, but this seemingly simple function can be extremely expensive to implement in practice.

\subsection{Decoding Algorithm}
\begin{figure*} [ht]
  \vspace{-0.3cm}
 \centerline{ \includegraphics[scale=0.6]{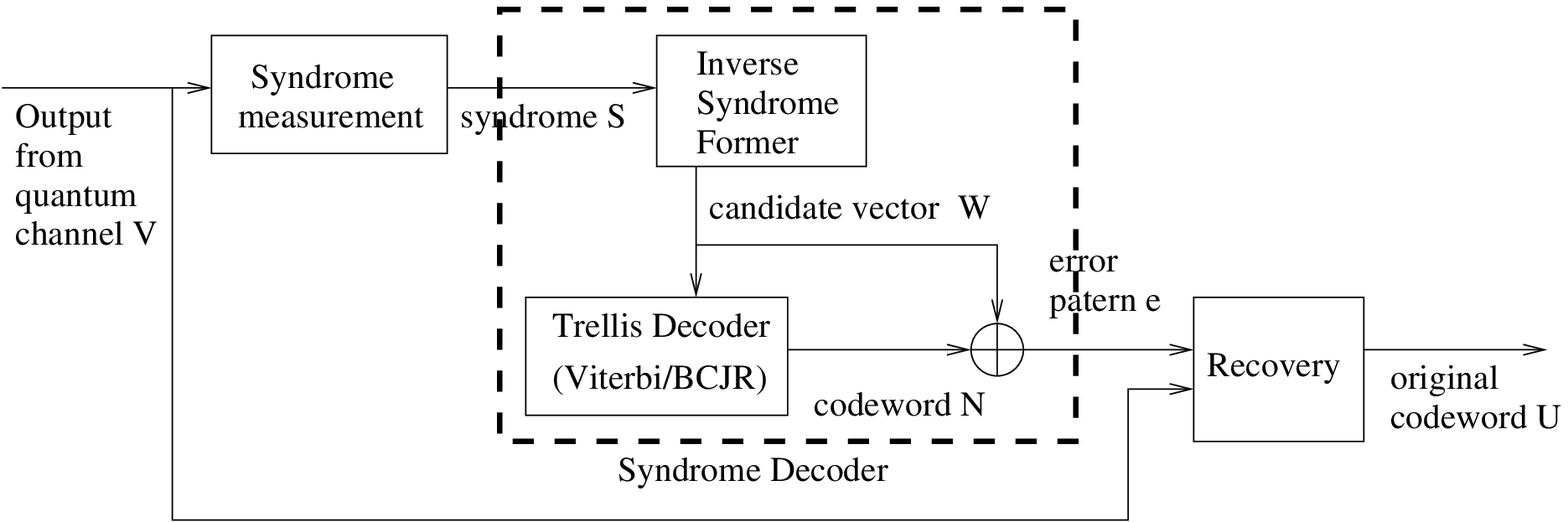}}
\vspace{-0.2cm}
  \caption{Diagram of general quantum decoding, composed of two stages: decoding error patterns, recovery.}
  \label{fig:system_model}
\vspace{-0.4cm}
\end{figure*}

Stabilizer codes in quantum coding are analogous to linear codes in classical coding, and may be tackled through a similar coset representation and syndrome decoding.
The propose decoder here is not only a {\it conceptual} quantum version of the syndrome decoder, but also a {\it practical} mechanism that provides a systematic and concrete way and practical procedure to realize the syndrome decoder with manageable complexity.  The proposed quantum syndrome decoding, which is a low-complexity ML decoding procedure, is depicted in Fig. \ref{fig:system_model}. Below we explain why, how, and how efficiently the proposed algorithm works. We start by  transforming  the stabilizer
generators to a binary polynomial form, and find its equivalent
transfer polynomial. We then detail the step-by-step approach in Fig. \ref{fig:system_model}, and demonstrate examples to show how to decode without (exhaustive) lookup tables.

Here are a few notations used in the discussion. A quantum convolutional code may be described by its \textit{semi-infinite stabilizer matrix $M$}, or   \textit{stabilizer polynomial $S(D)$}.
The proposed decoding procedure involves the derivation of the equivalent
$\mathbb{F}_2$ or $\mathbb{F}_4$ classical code from the
stabilizer polynomial $S(D)$, which are termed the \textit{equivalent}
$\mathbb{F}_2$ or $\mathbb{F}_4$) \textit{transfer polynomial}, and denoted by
$H_b(D)$ or $H_q(D)$, respectively. The decoding procedure also makes heavy use of useful concepts in classical coding: {\it syndrome former} (SF) and {\it inverse syndrome former} (ISF). For a given $(n,k)$ linear code and hence a deterministic standard array, the SF finds the syndrome associated with a given sequence (an error pattern), and the ISF finds {\it one} error pattern -- any one of the $2^k$ possibilities -- that associates with a given syndrome.

\vspace{0.2cm}
\noindent{\bf Decoding Algorithm:}

The ``syndrome decoder'' block (dashed box) in Fig. \ref{fig:system_model} illustrates the proposed quantum syndrome decoder, whose input is a binary syndrome sequence obtained from the syndrome measurement.
\begin{enumerate}
  \item Represent the original quantum convolutional code using its
  equivalent $\mathbb{F}_2$ transfer polynomial, which leads to an equivalent classical convolutional code.
  \item Perform syndrome decoding on the equivalent classical
  convolutional code, through the following two substeps:
\begin{enumerate}
  \item Compute the $\mathbb{F}_2$ ISF from
  the equivalent $\mathbb{F}_2$ transfer polynomial. Use the ISF to locate an
  arbitrary candidate vector in the coset corresponding to the given
  syndrome.
  \item Compute the $\mathbb{F}_2$ generator polynomial $G_b(D)$ from
  the equivalent $\mathbb{F}_2$ transfer polynomial. From the generator polynomial, implement a conventional trellis decoder (such as the Viterbi algorithm or the BCJR algorithm), and use it to decode the candidate vector into a valid codeword. Exclusive-OR (XOR) the candidate vector and the valid codeword to get the error pattern.
\end{enumerate}
\end{enumerate}
The output of the decoder is a $2n$-bit binary sequence, which can be re-assembled to an $n$-qubit 
Pauli-formed error pattern.
This error pattern is then fed into the ``recovery'' block in Fig. \ref{fig:system_model} and  counter-acted to recover the original quantum states.

\vspace{0.2cm}
\noindent{\bf{Step 1: Deriving equivalent  $\mathbb{F}_2$  transfer polynomial}}

Let $S(D)$ be the stabilizer polynomial of a quantum convolutional code.
Recall that a stabilizer matrix can be described either in the
binary $\mathbb{F}_2$ form, $S = (P|Q)$, where $P$ and $Q$ are
symplectically orthogonal, or in a quaternary $\mathbb{F}_4$ form,
where Pauli operators $\{{\bf I, Y, X, Z}\}$  are represented by  $\mathbb{F}_4$ elements $\{ 0, 1, \omega, \omega^2=\bar{\omega}\}$, respectively, conforming to the zero trace inner product ($\omega$ is a primary element in
$\mathbb{F}_4$). A stabilizer code in $\mathbb{F}_4$ form takes a
similar flavor as an $\mathbb{F}_4$ classic code, but the
syndrome takes a {\it binary}, rather than quaternary, value. Let
$M_j=(m_{1j},m_{2j})$ be a binary row in $(P|Q)$,
and let $E=(e_1,e_2)$ be a binary error vector. Without loss of
generality, re-write $E$ and $M_j$ in $\mathbb{F}_4$ as: $E = e_1 +
e_2 \omega$, $M_j = m_{1j} + m_{2j} \omega$. The syndrome
corresponding to $E$ is computed as\vspace{-0.2cm}
%
\begin{equation}
  s_j(E) = e_1\cdot m_2^T + e_2\cdot m_1^T.  \label{eqn:syndrome}
\end{equation}
An immediate and important implication of (\ref{eqn:syndrome}) is

\vspace{0.2cm}
\noindent {\bf Lemma 1:} \cite{bib:QC_ISIT}   An equivalent classical linear code to the quantum convolutional
code with stabilizer generator $S(D) = ( P(D) | Q(D) )$ has a binary
($\mathbb{F}_2$) transfer polynomial: $H_b(D) = P(D^2) + D Q(D^2)$. Here equivalent is meant in terms of decoding capability promised by the code (disregarding possible degeneracy).

\vspace{0.2cm}
\noindent{\bf{Step 2: Syndrome decoding for convolutional codes}}

The problem now boils down to performing syndrome decoding for the equivalent classical
convolutional codes.

For an ($n,k$) linear code, the function of the SF is to map a length-$n$
vector to a length-$(n\!-\!k)$ syndrome.
The ISF performs the opposite function of SF by locating a
length-$n$ vector in the coset associated with a given syndrome.
Since each coset contains $2^k$ vectors, there are $2^k$ possible outputs an ISF may produce, and
the resultant vector can be any of the $2^k$ vectors depending on the specific inverse syndrome former being implemented. The choice of the ISF will not affect the end-result of the proposed syndrome decoder.
Efficient and systematic ways exist to implement SF and ISF, the simplest of which is through linear transformation or linear circuits. For a linear block code, the parity check matrix can serve the role of SF, and the left inverse of the parity check matrix can serve as a matching ISF.




The proposed syndrome decoder makes essential use of the functions
of SF and ISF. The ``syndrome measurement'' block in Fig.
\ref{fig:system_model} generates a measure of the syndrome, and the
``syndrome decoder'' block uses a matching ISF  to acquire a valid
vector in the coset associated with that syndrome, and subsequently
locates the coset leader after trellis decoding and an XOR
operation. Finally, the odd-positioned bits and the even-positioned
bit in the  coset leader are paired up to obtain the tensor product
form $\mathbf{I, X, Y, Z}$ of the error pattern, which will then be
applied to the received quantum sequence (the ``recovery'' block in
Fig.\ref{fig:system_model}, a well-established quantum mechanics
procedure) to get the correct codeword.

\vspace{0.2cm}
\noindent {\bf Lemma 2:} The proposed syndrome decoding approach in Fig.\ref{fig:system_model} is valid.

\vspace{0.1cm}
\noindent{\it Proof:} The validity of this decoder is
warranted by the fact that the error pattern/coset leader is the
minimum-weight vector inside each coset.
Suppose that the transmitter sends a valid binary codeword
$U$. The discrete memoryless channel  adds some recoverable binary noise $e$ to $U$ to yield $V$. That is, $ V=U \oplus e$, and hence $e= U \oplus V$, where $\oplus$ denotes binary addition (XOR). The quantum receiver
collects $V$, but instead of having a direct measurement of $V$, it gets a measurement of the syndrome $S$. We show the the combination of ISF and trellis decoding in Fig. \ref{fig:system_model} successfully deduce $e$.

Inverse syndrome former maps $S$ to an arbitrary vector $W$ in that coset. In the standard array, $U$ locates in the first row
(corresponding to the zero syndrome, $V$, $e$ and $W$ all locate in
the same row corresponding to the syndrome $S$, $e$ is the
coset leader (because it is a recoverable error patten), and  $V$ locates in the same column as $U$ (because $U\oplus e=V$). Hence, $W$ is in fact some codeword corrupted by error $e$, namely,  there exist some valid codeword $N$ such that $N\oplus e =W$.

The vector $W$ at the output of the ISF is fed to the ML trellis decoder.
If the code can support this channel, then with a probability approaching 1, the trellis decoder is able  to deduce the valid codeword $N$ from $W$.  Since $e = U \oplus V = N \oplus W$, XORing $N$ (output from trellis decoder)
and $W$ (output from ISF) thus produces the target error pattern $e$. $\square$


%

\subsection{Realization of Important Modules in Decoder}

As discussed before, the proposed syndrome decoder in Fig. \ref{fig:system_model} involves two key modules, the ISF and a conventional trellis decoder for the equivalent classical convolutional code. The latter is a well-established procedure, given a generator polynomial. Lemma 1 establishes the relation between a stabilizer polynomial $S(D)$ of a quantum convolutional code and the transfer polynomial $H_b(D)$ of its equivalent classical linear convolutional code, below we focus the discussion on the derivation of the ISF and the generator polynomial for a given $H_b(D)$.


There exists intrinsic connection between $\mathbb{F}_2$ generator polynomial, transfer polynomial, SF and
ISF. Specifically, consider a
binary $k/n$ convolutional code with transfer polynomial $H_b(D)$. The transpose of the transfer polynomial, $(H_b)^T$, which has
dimension $n \times (n-k)$ can serve as an SF. The generator
polynomial can be implemented using a linear sequential circuit
specified by $k \times n$ transfer function $G_b$ with rank $k$ satisfying
\begin{equation}\label{equ:SF2G}
    G_b \times (H_b)^T = 0_k,
\end{equation}
where $0_k$ is a $k$-by-$(n-k)$ all-zero matrix. The constraint in
(\ref{equ:SF2G}) guarantees that all the valid codewords are
associated with the length-$(n-k)$ all-zero syndrome $0_{n-k}$ and
that  a set of length-$n$ codewords/vectors have the same syndrome if and
only if they belong to the same coset.

Inverse syndrome former can be obtained by taking the
left inverse of the syndrome former, and is therefore denoted by $(H_b^{-1})^T$:
\begin{equation}\label{equ:SF2ISF}
    (H_b^{-1})^T H_b^T = I_{n-k},
\end{equation}
where $I_{n-k}$ is an identity matrix with rank $n\!-\!k$.
The inverse syndrome former may be derived using the partial Gaussian
elimination or  the Moore-Penrose matrix inverse method.
The latter states
that the left inverse of a rectangular matrix $B$, denoted as
$B^+$, takes the form of $B^+ = (A^T \cdot A)^{-1} \cdot A^T$. Although
the Moore-Penrose matrix inverse was originally derived in real-valued matrix, it can be easily extended to binary matrix by with module-2
computation.

\vspace{0.2cm}
{\bf Example 1:} This example illustrates how to obtain the ISF and the
generator polynomials from a given SF. Supposed the given SF takes of the form of
\begin{equation}
H_b^T = \left(
  \begin{array}{c}
    1 + D^2 \\
    1 + D + D^2
  \end{array}
\right), \nonumber
\end{equation}
which corresponds to a rate 1/2 (classical) recursive convolutional code.
Using the Moore-Penrose procedure, a possible ISF and generator polynomial may be obtained:
\begin{eqnarray}
  \mbox{ISF}:&\ &  (H_b^{-1})^T = \left(
                        \begin{array}{cc}
                          1+D & D
                        \end{array}
                      \right);\\
  \mbox{GP}:&\ &  G_b = \left(
               \begin{array}{cc}
                 1 & \frac{1+D^2}{1+D+D^2}
               \end{array}
             \right).
\end{eqnarray}

For a given SF, matching  ISFs and generator polynomials are not unique. For
instance, in the previous example, another possible choice for ISF and generator polynomial are:
\begin{eqnarray}
\mbox{ISF}: &\  & (H_b^{-1})^T = \left(
                        \begin{array}{cc}
                          1 & \frac{1+D^2}{1+D+D^2}
                        \end{array}
                      \right);\\
\mbox{GP}:&\ & G_b = \left(
               \begin{array}{cc}
                 1+D+D^2 & 1+D^2
               \end{array}
             \right).
\end{eqnarray}
It is desirable to choose a non-catastrophic generator polynomial, and an inverse syndrome former with a low complexity.

\vspace{0.2cm}
\noindent{\bf Example 2:}
This example illustrates the entire decoding procedure.
Consider the quantum convolutional code in \cite{bib:TailbitingCC},
which has a semi-infinite stabilizer matrix (in Pauli operator form):
\begin{equation}\label{example_pauli_stabilizer}
M = \left(
  \begin{array}{cccccccccc}
    X & X & X & X & Z & Y & & & &  \\
    Z & Z & Z & Z & Y & X & & & &  \\
      &   &   & X & X & X & X & Z & Y & \\
      &   &   & Z & Z & Z & Z & Y & X & \\
      &   &   &   &   &   &   &   &   & \ddots \\
  \end{array}
\right).
\end{equation}
It represents a $[n,k,m]=[3,1,1]$ quantum convolutional code with rate 1/3.


Following the proposed syndrome decoding algorithm, the first step is to derive
 the equivalent $\mathbb{F}_2$ transfer
polynomials:
\begin{equation}\label{example_SF}
H_b(D) = \left(
  \begin{array}{ccc}
    1+D^2 & 1+D^3 & 1+D^2+D^3 \\
    D+D^3 & D+D^2+D^3 & D+D^2 \\
  \end{array}
\right),
\end{equation}
whose matrix transpose, $H_b(D)^T$, represents the syndrome former of  a ($3, 1, 2$) $\mathbb{F}_2$ (classical) convolutional code.

The next step is to derive the ISF $(H_b^T(D))^{-1}$ and the generator polynomial
$G_b(D)$ matched to the given SF $H_b(D)^T$ (all in $\mathbb{F}_2$). Following
the Moore-Penrose procedure, we get
\begin{eqnarray}\label{example_ISF}
\mbox{ISF}: & \ &(H_b^T(D))^{-1} = \left(
  \begin{array}{ccc}
    \frac{1}{D+D^3} & \frac{1}{D+D^2+D^3} & 0 \\
    \frac{1}{D^2+D^3} & \frac{1}{D^2+D^3+D^4} & 0 \\
  \end{array}
\right),\\
\mbox{GP}: & \ &
G_b(D) = \left(
  \begin{array}{ccc}
    D^2 & 1+D^2 & 1+D^2
  \end{array}
\right). \label{example_G}
\end{eqnarray}

\begin{figure} [ht]
\vspace{-0.2cm}
  \includegraphics[width=3.4in]{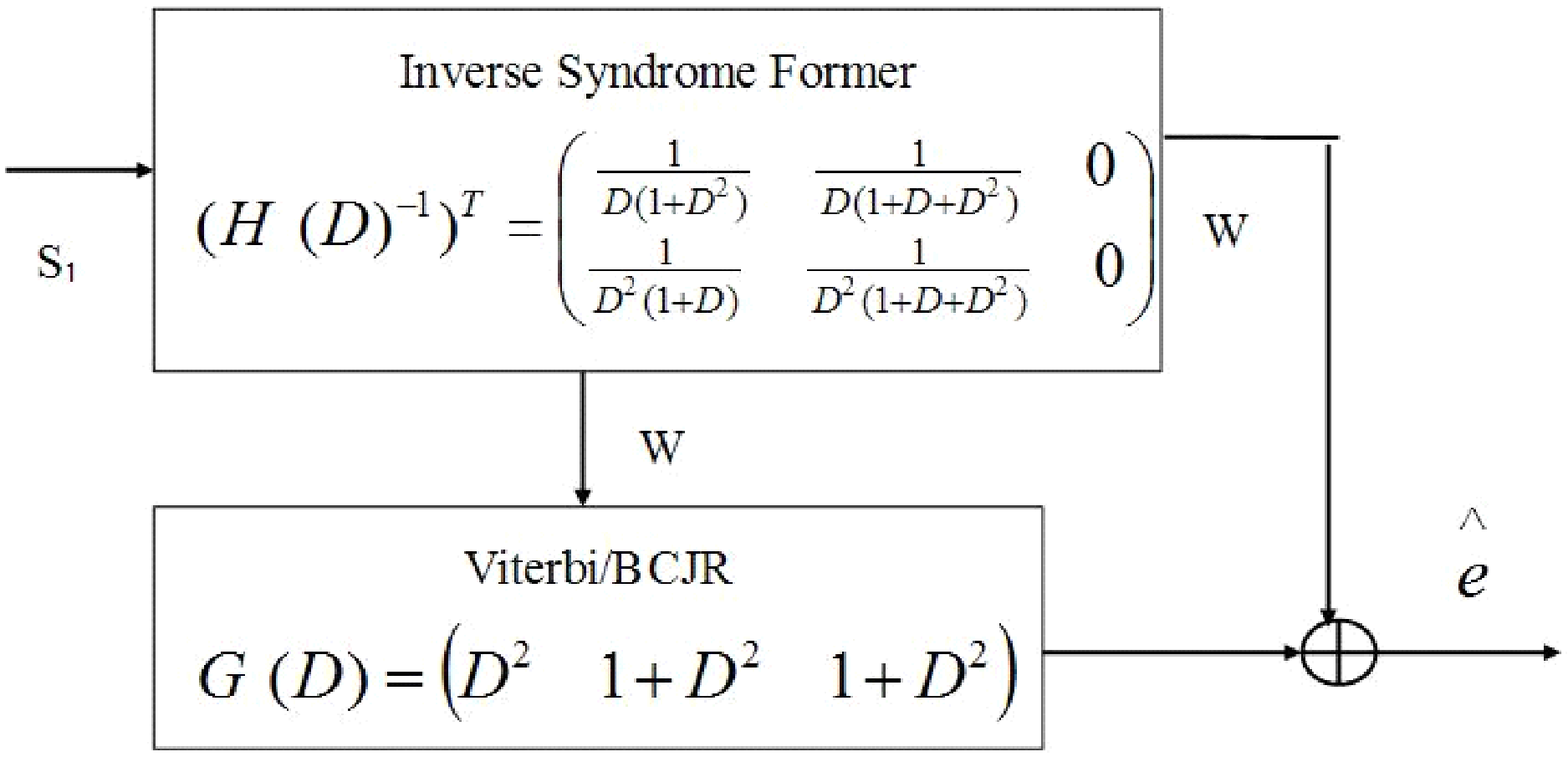}
\vspace{-0.2cm}
  \caption{Practical decoder of quantum convolutional code (example 2).}
  \label{fig:QCC_decoder}
\vspace{-0.4cm}
\end{figure}

The resultant decoder diagram is shown in Fig. \ref{fig:QCC_decoder}.
 The binary syndrome sequence $S_1$ is separated
into two streams, and feeds into the inverse syndrome former, which, in this example, is like a recursive convolutional encoder with two streams in and three streams out. The
output streams assemble to a sequence $W$, passes through a BCJR decoder, and generates a
sequence which is a valid codeword. The difference between the codeword
sequence and $W$ is the error pattern sequence $\hat{e}$. $\hat{e}$ is a binary
sequence, and will be divided into 2 parts and packed back to Pauli
operators accordingly.

\subsection{Decoder Implementation in $\mathbb{ F}_4$}

The proposed decoder is applicable to general quantum
convolutional codes on a variety of quantum channel models characterized by
memoryless independent Pauli operators.
The previous discussion and examples are centered around $\mathbb{F}_2$ implementation. When the quantum convolutional code is derived from an
$\mathbb{F}_4$-linear codes, such as those proposed in \cite{bib:TailbitingCC},
 one may either find its equivalent binary convolutional code as discussed before, or explore syndrome decoding on its equivalent $\mathbb{F}_4$
counterpart. The latter is possible, because the syndromes, which are typically $(n\!-\!k)$-bit binary sequences, may also be represented by a length $(n\!-\!k)/2$ $\mathbb{F}_4$ sequences in this example.

\vspace{0.2cm}
\noindent {\bf Example 3:}
Consider the rate-1/3 [$3,1,1$] quantum convolutional
code from Example 2, whose equivalent $\mathbb{F}_4$
convolutional code has a $\mathbb{F}_4$ transfer polynomial $H_q(D)$
\begin{equation}\label{example_generator}
   H_q(D) = (1+D, \ \ \ 1+\omega D, \ \ \ 1+\bar{\omega} D).
\end{equation}
The $\mathbb{F}_4$ syndrome former is given by $H_q(D)^T$.

The ISF and the generator polynomial may also be derived in $\mathbb{F}_4$ form:
\begin{eqnarray}
\mbox{ISF}:  & \ & (H_q^{-1}(D))^T = \left(
                        \begin{array}{ccc}
                          1, & 1, & 1
                        \end{array}
                      \right),\\
\mbox{GP}: &\ &  G_q(D) = \left(
               \begin{array}{ccc}
                 0, & 1+\bar{\omega}D, & 1+\omega D \\
                 1+\omega D, & 1+D, & 0
               \end{array}
             \right).
\end{eqnarray}
Clearly, the trellis decoder (Viterbi/BCJR) in Fig. \ref{fig:system_model} can be implemented based on $\mathbb{F}_4$ generator polynomial. The entire syndrome decoding is carried out over a $(n,k,m)=(3, 2, 1)$ classical
$\mathbb{F}_4$ convolutional code.


\section{Simulation Results}
\label{sec:Simulation}

We demonstrate the proposed algorithm by simulation on a memoryless
bi-polar Pauli error channel, where ${\bf X}$ and ${\bf Z}$ occur at
probability $p-p^2$, and ${\bf Y}$ occurs at probability $p^2$.

\begin{figure}[htbf]
  \vspace{-0.3cm}
  \centering{\includegraphics[width=3.4in]{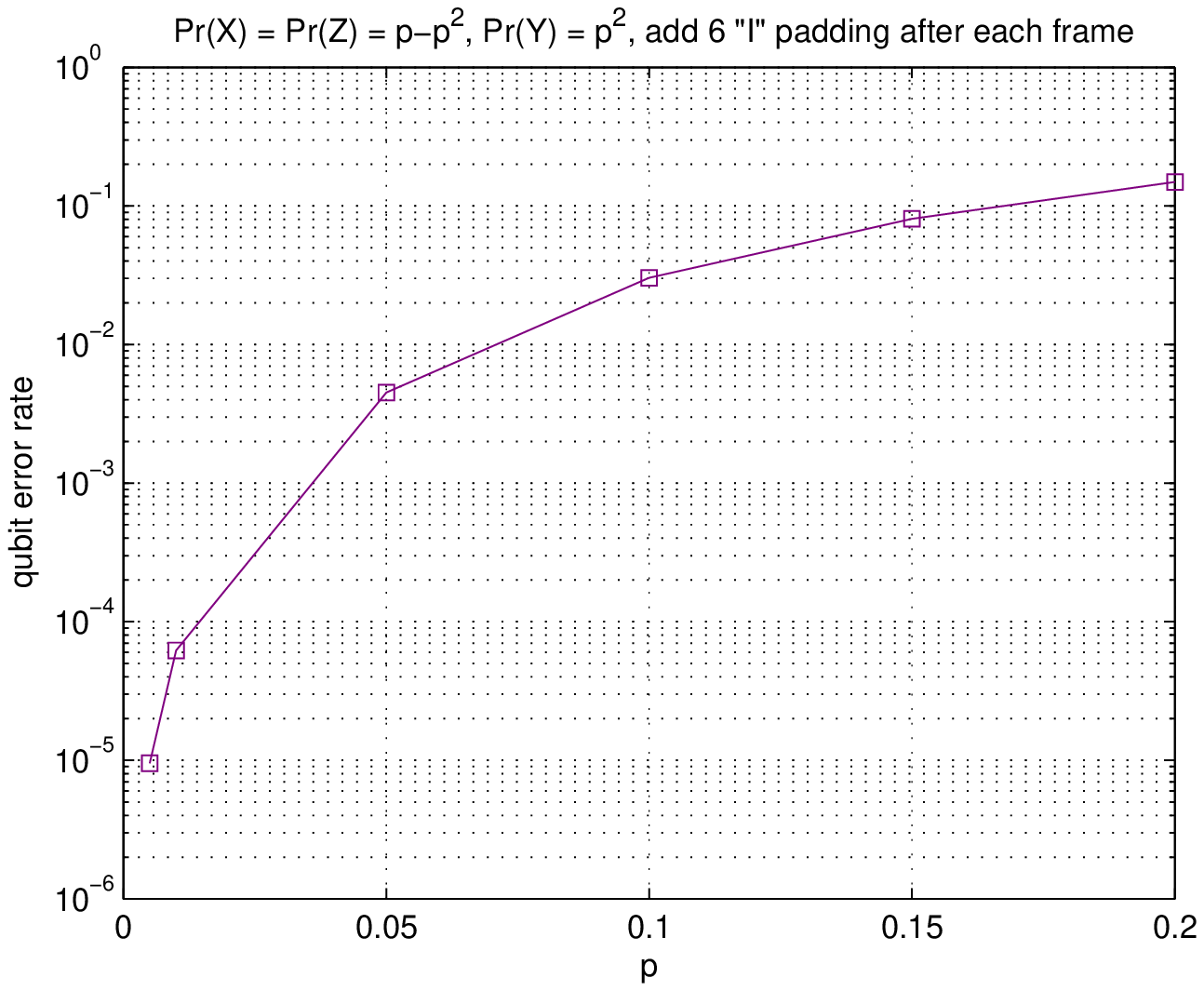}\vspace{-0.3cm}}
  \caption{Simulation results of quantum convolutional codes.}
\vspace{-0.3cm}
  \label{fig:performance}
\vspace{-0.4cm}
\end{figure}

Notice that the ISF is essentially a recursive convolutional
code, padding bits are necessary to generate correct $W$ for
simulations. 
For an [$n,k,m$] quantum convolutional code, the padding sequence should be
 a length-$(n+nm)$ $\mathbf{I}$ sequence.
We simulated the [3,1,1] quantum convolutional code at a frame length
of 900 qubits and tested 10000 frames. This is like a $[900, 300]$ stabilizer block with an additional 6-bit padding (overhead), so the exact code
rate is $300/(900+6)\approx 1/3$. The resultant performance curve is shown in Fig. \ref{fig:performance}.


\section{Conclusion}
\label{sec:conclusion}

We have developed a new syndrome decoding approach, which is
applicable to a general stabilizer convolutional code on a general
Pauli-error channel model. Comparing with the existing  quantum
Viterbi decoder and its simplified version, the new algorithm enjoys
significantly lower complexity (no lookup table, and one-time
Viterbi) without scarifying ML performance. The decoding algorithm
roots to the classical syndrome decoding theory, and boasts provenly
optimal performance. Using this algorithm, we simulated and
presented the first performance curve for a general quantum
convolutional code.


%

\end{document}